\documentclass[reprint,
superscriptaddress,
amsmath,amssymb,
aps,prb,
floatfix
]{revtex4-2}

\usepackage{graphics}
\usepackage{hyperref}
\usepackage{amssymb}
\usepackage{amsmath,mathtools}
\usepackage[framemethod=TikZ]{mdframed}
\usepackage{xcolor,colortbl}

\newcommand{\ve}[1]{\boldsymbol{#1}}

\begin{document}
\title{Photoemission orbital tomography for excitons in organic molecules}

\author{C. S. Kern}
\author{A. Windischbacher}
\author{P. Puschnig}
\email[E-mail address: ]{peter.puschnig@uni-graz.at}
\affiliation{Institute of Physics, NAWI Graz, University of Graz, 8010 Graz, Austria}

\begin{abstract}
Driven by recent developments in time-resolved photoemission spectroscopy, we extend the successful method of photoemission orbital tomography (POT) to excitons. Our theory retains the intuitive orbital picture of POT, while respecting both the entangled character of the exciton wave function and the energy conservation in the photoemission process. Analyzing results from three organic molecules, we classify generic exciton structures and give a simple interpretation in terms of natural transition orbitals. We validate our findings by directly simulating pump-probe experiments with time-dependent density functional theory.
\end{abstract}

\date{\today}
\maketitle 

\section{Introduction}
In the past decade, photoemission orbital tomography (POT)~\cite{Puschnig2009a,Dauth2011,Nguyen2015,Woodruff2016,Puschnig2017,Kliuiev2019} has emerged as a powerful technique that relates the measured photoemission angular distribution (PAD) from oriented films of organic molecules with the orbitals from which the electron has been emitted. This direct connection arises from modeling photoelectrons by plane waves. While this plane wave assumption has been debated~\cite{Bradshaw2015,Egger2018} and, in fact, demonstrated to be insufficient in certain circumstances~\cite{Dauth2016a,Metzger2020}, POT has led to valuable insights, for instance, into the hybridization between organic layers and the substrate~\cite{Ziroff2010,Zamborlini2017,Yang2022}, the geometry of adsorbate layers~\cite{Graus2016,Kliuiev2019,Hurdax2022}, the nature of reaction products~\cite{Haags2020} or real space images of orbitals~\cite{Luftner2013,Weiss2015,Kliuiev2016a,Graus2019,Jansen2020}. Particularly the latter aspect has also stimulated discussions on how to build a formal bridge between quantum mechanical wave functions and the experimentally observed momentum space distributions~\cite{Truhlar2019,Krylov2020}.

Despite these numerous achievements, until very recently, POT could only be applied to study occupied molecular orbitals by photoexciting electrons from the ground state. With the advent of laser high-harmonic generation and free-electron lasers, it has become possible to study also the dynamics of excited states in time- and angle-resolved photoemission spectroscopy (trARPES) experiments. On the one hand, band structure movies of crystalline solids have shown the temporal evolution of the electron system over the complete Brillouin zone~\cite{Rohwer2011,Eich2017,Nicholson2018}. 
On the other hand, for molecular systems, optically excited states, involving transitions from HOMO to LUMO, have recently been observed with trARPES on their intrinsic temporal and spatial scales~\cite{Wallauer2020,Baumgartner2022,Neef2023}.
In more complex systems however, the simple HOMO-LUMO picture breaks down and excitons may involve multiple transitions as, for instance, observed in van der Waals heterostructures~\cite{Wallauer2016,Bertoni2016,Madeo2020,Wallauer2021,Hagel2021,Bange2023a,Bange2023b} and defects therein~\cite{RefaelyAbramson2018,Mitterreiter2021,Hoetger2023}. In organic semiconductor crystals, the multi-orbital nature of excitons is crucial~\cite{Bennecke2023} and also relevant for understanding singlet fission~\cite{Neef2023}. Thus, an exciton must be generally treated as an entangled state composed of multiple electron-hole transitions for which a theoretical foundation of POT is still lacking.

The aim of this work is to fill this gap and establish a consistent framework that allows us to interpret measured PAD maps from excitons. Specifically, we assume that the exciton wave function is represented in a product basis of valence and conduction states, as typically done when solving the electron-hole Bethe-Salpeter equation (BSE)~\cite{Rohlfing2000} or Casida's equation in time-dependent density functional theory (TDDFT)~\cite{Casida1995,Onida2002}. Expanding the concept of Dyson orbitals~\cite{Pickup1977,Krylov2020,Ortiz2020} to excited states, we arrive at the result that the PAD can be interpreted as the Fourier-transformed coherent sum of the electronic part of the exciton wave function. These relations, as well as the unexpected consequences of the photohole's state for the measured kinetic energy spectrum, are illustrated for generic cases of exciton compositions in a series of organic molecules in the gas phase. We further show how exciton photoemission can be interpreted in terms of the established concept of natural transition orbitals (NTOs)~\cite{Martin2003} and, underpinning our findings, the PAD is also simulated directly by means of a TDDFT approach where no assumptions on the final state are made whatsoever.

\section{Theory}
Bound electron-hole pairs, excitons, are the fundamental optical excitations for energies below the band gap in molecules and non-metallic solids.
For such correlated electron-hole pairs, we assume that the wave function of the $m$-th exciton, with excitation energy $\Omega_m$, can be expanded in the single-particle electron $\{\chi_c(\ve{r}_e)\}$ and hole basis $\{\phi_v(\ve{r}_h)\}$ as
\begin{equation}
\label{eq:quasiparticle}
\psi_m(\ve{r}_h, \ve{r}_e) = \sum_{v,c} X^{(m)}_{vc} \phi_v^*(\ve{r}_h) \chi_c(\ve{r}_e).
\end{equation}
Here, the sum runs over all pairs of valence and conduction states $\{v,c\}$, respectively, and $X^{(m)}_{v c}$ is the transition density matrix that describes the character of the exciton.
Note that here and in the following derivations, we use the Tamm-Dancoff approximation~\cite{Benedict1989} for better readability. In the general case and in our calculations, however, we also consider de-excitations.

\subsection{Photoemission from Excitons}
Our goal is to find a consistent expression that connects the exciton wave function as defined in Eq.~\ref{eq:quasiparticle} with measured photoemission momentum maps. In the spirit of POT, we describe the photoelectron probability with Fermi's golden rule as the transition from an initial $N$-particle state $\Psi_\mathrm{i}^N$ to a final state $\Psi_\mathrm{f}^N$, triggered by the photon field $\ve{A}$ with energy $\omega$. We couple this classical field to the electrons' momenta $\ve{P}$ in the dipole approximation and use the Coulomb gauge as well as Hartree atomic units unless stated otherwise.
Denoting the energy of the states $\Psi_\mathrm{i}^N$ and $\Psi_\mathrm{f}^N$ with $E_\mathrm{i}$ and $E_\mathrm{f}$ respectively, the photoelectron probability is
\begin{equation}
\label{eq:goldenrule}
W_{\mathrm{i} \rightarrow \mathrm{f}} = 2 \pi \left| \left\langle \Psi_\mathrm{f}^N \right| \ve{A} \ve{P} \left| \Psi_\mathrm{i}^N \right\rangle \right|^2
 \delta\left(\omega + E_\mathrm{i} - E_\mathrm{f} \right).
\end{equation}
In contrast to earlier work on photoemission from the electronic ground state  $\Psi_\mathrm{i,0}^N$, now the initial state is given by the  the $m$-th exciton which can also be expressed in a second quantization formulation as
\begin{equation}
\label{eq:initialstate}
\left| \Psi_{\mathrm{i},m}^N \right\rangle = \sum_{v,c} X^{(m)}_{vc}
a^\dagger_c a_v 
\left|\Psi_\mathrm{i,0}^N \right\rangle.
\end{equation}
Here, $a_v$ and $a^\dagger_c$ create a hole and an electron in state $v$ and $c$, respectively. Its energy $E_{\mathrm{i},m}^N$ is the sum of the ground state energy $E_{\mathrm{i},0}^N$ and the excitation energy $\Omega_m$.

For the final state $\Psi_\mathrm{f}^N$, one commonly assumes the sudden approximation, in which the correlation between the emitted electron $\gamma_{\ve{k}}$ and the remaining system can be neglected~\cite{Damascelli2004}, and $\Psi_\mathrm{f}^N$ can be written as the anti-symmetrized product of the $(N-1)$ electron state $\Psi_{\mathrm{f},j}^{N-1}$ and the photoemitted electron with momentum $\ve k$ in state $\gamma_{\ve{k}}$:
\begin{equation}
  \Psi_{\mathrm{f},j,\ve{k}}^{N} = \mathcal{A} 
  \Psi_{\mathrm{f},j}^{N-1} \gamma_{\ve{k}}.
\end{equation}
Like the initial state, $\Psi_{\mathrm{f},j}^{N-1}$ may be expressed in Fock space, i.e.~as the $N$-electron ground state from which the $j$-th electron has been removed:
\begin{equation}
  \label{eq:finalstate2}
  \left| \Psi_{\mathrm{f},j}^{N-1} \right\rangle = a_j
  \left|\Psi_{\mathrm i,0}^N \right\rangle.
\end{equation}
Under these assumptions, we can identify the total energy of this final state  as the sum of $E_{\mathrm{f},j}^{N-1}$ and the photoelectron's kinetic energy, $E_\mathrm{kin} = k^2 / 2$. The energy conservation from Eq.~\ref{eq:goldenrule} then demands \cite{Weinelt2004}
\begin{eqnarray}
E_\mathrm{kin} = \omega - (E_{\mathrm{f},j}^{N-1} - E_{\mathrm{i},0}^N) + \Omega_m = \omega - \varepsilon_j + \Omega_m,
\end{eqnarray}
where we have introduced the $j$-th ionization potential $\varepsilon_j$ as the energy difference between the $j$-th excited state of the $(N-1)$-electron system and the $N$ electron ground state. 
In taking the overlap between the two wave functions for the $N$-electron and the $(N-1)$-electron system, we utilize the Dyson orbital for electron detachment $D_{j,m}$, in the usual way~\cite{Pickup1977,Oana2007,Krylov2020}, with the only modification that in our case the Dyson amplitudes have to be spanned over both the basis sets $\{\varphi_{v'}\}$ and $\{\chi_{c'}\}$:
\begin{align}
 \label{eq:dyson2}
 D_{j,m}(\mathbf r) =
 &\sum_{v'} \left\langle \Psi_{\mathrm{i},m}^N
 \middle| a_{v'}^{\dagger} \middle|
 \Psi_{\mathrm{f},j}^{N-1} \right\rangle \phi_{v'} (\mathbf r)+ \nonumber \\
 +&\sum_{c'} \left\langle \Psi_{\mathrm{i},m}^N
 \middle| a_{c'}^{\dagger} \middle|
 \Psi_{\mathrm{f},j}^{N-1} \right\rangle \chi_{c'} (\mathbf r).
\end{align}
It is accepted that Dyson orbitals represent the most appropriate way to describe photoemission in a single-orbital picture~\cite{Oana2007,Dauth2014,Gozem2015,Truhlar2019,Krylov2020}, however, their computation from correlated wave functions in a multi-reference framework~\cite{Oana2009,Vidal2020} is often not feasible. Therefore, and with weakly-correlated systems in mind, we approximate $\Psi_\mathrm{i,0}^N$ by a single Slater determinant.
Inserting the $N$-electron wave function, Eq.~\ref{eq:initialstate}, and the $(N-1)$-electron wave function, Eq.~\ref{eq:finalstate2}, into the expression for the Dyson orbital, we get
\begin{align}
  D_{j,m}(\mathbf r) =
 &\sum_{v'} \sum_{v,c} X_{v c}^{(m)} \left\langle \Psi_\mathrm{i,0}^N 
 \middle| a_v^{\dagger} a_c a_{v'}^{\dagger} a_j \middle|
 \Psi_{\mathrm i,0}^N \right\rangle \phi_{v'} (\mathbf r)+\nonumber \\
 +&\sum_{c'} \sum_{v,c} X_{v c}^{(m)} \left\langle \Psi_\mathrm{i,0}^N 
 \middle| a_v^{\dagger} a_c a_{c'}^{\dagger} a_j \middle|
 \Psi_{\mathrm i,0}^N \right\rangle \chi_{c'} (\mathbf r),
\end{align}
where all integrals in the sum over $v'$ vanish due to orthogonality. In the sum over $c'$, we get no contributions for $c \ne c'$ by the same argument and thus arrive at our final result for the $j$-th Dyson orbital (up to normalization constants):
\begin{align}
    \label{eq:coherent_sum}
     D_{j,m}(\mathbf r) &=
     \sum_{v,c} X_{v c}^{(m)} \left\langle \Psi_\mathrm{i,0}^N 
     \middle| a_v^{\dagger} a_c a_{c}^{\dagger} a_j \middle|
     \Psi_{\mathrm i,0}^N \right\rangle \chi_{c} (\mathbf r) = \nonumber \\
     &=
     \sum_{c} X_{j c}^{(m)} \left\langle \Psi_\mathrm{i,0}^N 
     \middle| a_j^{\dagger} a_j a_c a_{c}^{\dagger} \middle|
     \Psi_{\mathrm i,0}^N \right\rangle \chi_{c} (\mathbf r) = \nonumber \\
     &=
     \sum_{c} X_{j c}^{(m)} \chi_{c} (\mathbf r).
\end{align}
Note that exploiting the orthogonality relations between many-body wave functions in different states is possible here, since $\Psi_\mathrm{i,0}^N$ is represented by a single Slater determinant only. However, we remark that the above derivation could be extended to multi-configuration methods, albeit at the expense of an additional summation over configuration space in Eq.~\ref{eq:coherent_sum}.

With the help of the Dyson orbitals, we can avoid the explicit treatment of the $N-1$ passive electrons in the process and thereby reduce the matrix element of Eq.~\ref{eq:goldenrule} to an integral over a single coordinate only:
\begin{eqnarray}
\label{eq:me2}
\left\langle \Psi_{\mathrm{f},j}^N \right| \ve{A} \ve{P} \left| \Psi_{\mathrm{i},m}^N \right\rangle & \approx &
\ve{A} \int \mathrm{d}^3 r \; \overline{\gamma}_{\ve{k}}(\ve{r}) \, \ve{p} \, D_{j,m}(\ve{r}) \nonumber \\
& \propto & (\ve{A} \ve{k}) \, \mathcal{F}\left[ D_{j,m} \right] (\ve{k}).
\end{eqnarray}
In the second line, we make use of the plane wave approximation, $\gamma_{\ve{k}}(\ve{r}) \propto \mathrm{e}^{\mathrm i \ve{k} \ve{r}}$, that is inherent to POT~\cite{Puschnig2009a,Puschnig2017} and that naturally introduces
the Fourier transform of the Dyson orbital, modulated by a weakly angle-dependent polarization factor $\ve A \ve k$. Importantly, only the $j$-th row of the transition density matrix $X^{(m)}_{vc}$ contributes to the $j$-th Dyson orbital in Equation~\ref{eq:coherent_sum}, thereby fixing the hole position in the orbital $\phi_j$. Finally, the photoemission angular distribution arising from the $m$-th exciton is obtained by summing over all possible final state hole configurations as follows
\begin{eqnarray}
\label{eq:finalintensity}
I_m(\ve{k}) & \propto  & 
\left| \ve{A} \ve{k} \right|^2
\sum_{j} \left| \sum_{c} X^{(m)}_{j c}  \mathcal{F}\left[\chi_{c} \right] (\ve k)\right|^2 \nonumber \\
& \times &  \delta\left(\omega - E_\mathrm{kin} - \varepsilon_j + \Omega_m \right).
\end{eqnarray}
From this expression, which we refer to as ''exPOT'' (exciton POT) in the remainder of this work, we expect the photoemission signal from a general exciton to have contributions at multiple kinetic energies that are in concordance with the energy conservation and thus depend on the hole's position after electron detachment described by the ionization energy $\varepsilon_j$. At each allowed kinetic energy, momentum maps take the form of a Fourier transform of the \emph{coherent} sum over unoccupied states, weighted by the corresponding transition density matrix elements. 

\subsection{Formulation with Natural Transition Orbitals}
While the orbitals $\chi_c$ and the transition density matrix $X_{v c}$ appearing in the photoemission intensity expression for exPOT (Eq.~\ref{eq:finalintensity}) can be readily computed from a BSE or Casida calculation, physical intuition about the character of the exciton can be enhanced by introducing \emph{natural transition orbitals} (NTOs)~\cite{Martin2003,Krylov2020}.

Let us assume that in the exciton calculation there are $N_v$ occupied orbitals  $\phi_v$, and a number of $N_c$ unoccupied (or virtual) orbitals $\chi_c$. Then, the transition density matrix $X_{vc}$ is a matrix with $N_v$ rows and $N_c$ columns, whose singular value decomposition can be written in the following way
\begin{equation}
\label{eq:svd}
X = V \, \Lambda \, C^T.
\end{equation}
Here, $V$ and $C$ are quadratic matrices of sizes $N_v \times N_v$ and $N_c \times N_c$, respectively, and the rectangular $(N_v \times N_c)$-matrix $\Lambda$ has only non-vanishing elements $\lambda_1, \lambda_2,\dots \lambda_{N_v}$ in the diagonal. These singular values are ordered according to their magnitude, thus $\lambda_1 > \lambda_2 > \dots > \lambda_{N_v}$, and fulfill the normalization condition
\begin{equation}
\sum_{i=1}^{N_v} \lambda_i^2 = 1.
\end{equation}
Note that we have assumed that $N_v < N_c$ as is typically the case in the calculation of optically excited states. By making use of the transformations
\begin{eqnarray}
\widetilde{\phi}_\lambda & = & \sum_{v=1}^{N_v} V_{\lambda v}^T \phi_v \\
\widetilde{\chi}_\lambda & = & \sum_{c=1}^{N_c} C_{\lambda c}^T \chi_c,
\end{eqnarray}
we obtain a new set of orbitals, the NTOs $\widetilde{\phi}_\lambda$ and $\widetilde{\chi}_\lambda$, respectively, which can be used to express the exciton wave function in the electron-hole-basis (Eq.~\ref{eq:quasiparticle}):
\begin{equation}
\label{eq:NTOexciton}
\psi(\ve{r}_h, \ve{r}_e) = \sum_{\lambda=1}^{N_v}  \Lambda_{\lambda} \widetilde{\phi}_{\lambda}^*(\ve{r}_h) \widetilde{\chi}_{\lambda}(\ve{r}_e).
\end{equation}
Inserting Eq.~\ref{eq:svd} into Eq.~\ref{eq:finalintensity}---and by making use of the fact that the Fourier transform $\mathcal{F}$ is a linear operator---we can rewrite the exPOT formula for the photoemission intensity in the NTO basis as follows:
\begin{eqnarray}
\label{eq:nto_intensity}
I_m(\ve{k}) & \propto &
\left| \ve{A} \ve{k} \right|^2
\sum_{j} \left| \sum_{\lambda} V_{j \lambda} \Lambda_{\lambda} \mathcal{F}\left[\widetilde{\chi}_{\lambda} \right] (\ve k)\right|^2  \nonumber \\
& \times & \delta\left(\omega + E_\mathrm{kin} - \varepsilon_j + \Omega_m \right).
\end{eqnarray}
At first sight, it seems that we have not gained much: we have just replaced the summation over $c$ with the summation over $\lambda$ and replaced the prefactors. In practice, however, a given exciton is often characterized by just a few NTOs and one can easily control the accuracy of the exciton's representation in terms of NTOs by introducing a threshold for the $\Lambda_{\lambda}$. Moreover, it is our believe that NTOs are useful when dealing with excitons, since the character of the transition is encoded in just a few single-particle orbitals and with introducing Eq.~\ref{eq:nto_intensity}, we can assign physical meaning to these orbitals as actual observables of the excited-state photoemission experiment.

\subsection{Generic Exciton Structures}
\label{subsec:gen_ex_struct}
\begin{figure}[h!]
\begin{center}
\includegraphics[width=\columnwidth]{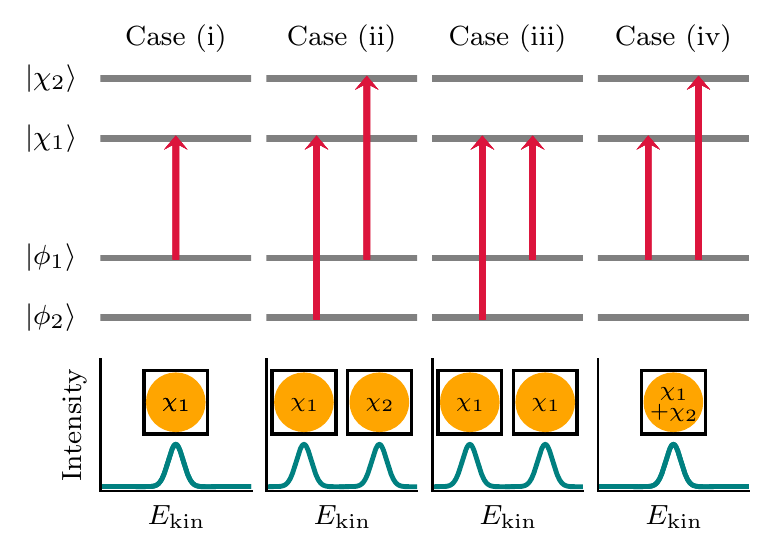}
\caption{Four prototypical exciton structures and the corresponding predictions of exPOT for the observed PAD maps as detailed in the text.}
\label{figure1}
\end{center}
\end{figure}

Before presenting our numerical results, we explain the implications of Eq.~\ref{eq:finalintensity} on the example of four prototypical exciton structures that are collected in Table~\ref{tab:fourcases} and schematically depicted in Figure~\ref{figure1}. For educational reasons, here only $N_v=2$ occupied and $N_c=2$ unoccupied orbitals are taken into account for setting up the transition density matrix such that all matrices are simple $2 \times 2$ matrices.
\begin{table}
\caption{Transition density matrices $X_{vc}$ as well as their singular value decompositions $X = V \Lambda C^T$ for the four simple exciton structures defined in Fig.~\ref{figure1}. Additionally, the exciton wave functions $\psi$ are given in terms of the NTOs $\widetilde{\phi}$ and $\widetilde{\chi}$, respectively. {\label{tab:fourcases}}}
\centering
\begin{tabular}{p{.6cm}p{1.35cm}p{1.9cm}p{2.05cm}p{2.05cm}}
\hline \hline
\\[-.5em]
   &  Case (i)  &  Case (ii)  & Case (iii)    &  Case (iv) \\
&&&&\\[-.7em]
\hline 
&&&&\\[-.5em]
$X=$ & 
$\left( \begin{array}{cc} 1 & 0 \\ 0 & 0 \end{array} \right)$ &
$\left( \begin{array}{cc} 0 & \frac{1}{\sqrt{2}} \\ \frac{1}{\sqrt{2}} & 0 \end{array} \right)$ &
$\left( \begin{array}{cc} \frac{1}{\sqrt{2}}  & 0 \\ \frac{1}{\sqrt{2}}  & 0 \end{array} \right)$ &
$\left( \begin{array}{cc} \frac{1}{\sqrt{2}}  & \frac{1}{\sqrt{2}}  \\ 0 & 0 \end{array} \right)$ \\
&&&&\\[-.8em]
$\Lambda=$ & 
$\left( \begin{array}{cc} 1 & 0 \\ 0 & 0 \end{array} \right)$ &
$\left( \begin{array}{cc} \frac{1}{\sqrt{2}} & 0 \\ 0 & \frac{1}{\sqrt{2}} \end{array} \right)$ &
$\left( \begin{array}{cc} 1 & 0 \\ 0 & 0 \end{array} \right)$ &
$\left( \begin{array}{cc} 1 & 0 \\ 0 & 0 \end{array} \right)$  \\
&&&&\\[-.8em]
$V=$ & 
$\left( \begin{array}{cc} 1 & 0 \\ 0 & 1 \end{array} \right)$ &
$\left( \begin{array}{cc} 0 & 1 \\ 1 & 0 \end{array} \right)$ &
$\left( \begin{array}{cc} \frac{1}{\sqrt{2}} & -\frac{1}{\sqrt{2}} \\ \frac{1}{\sqrt{2}} & \frac{1}{\sqrt{2}} \end{array} \right)$ &
$\left( \begin{array}{cc} 1 & 0 \\ 0 & 1 \end{array} \right)$  \\
&&&&\\[-.8em]
$C=$ & 
$\left( \begin{array}{cc} 1 & 0 \\ 0 & 1 \end{array} \right)$ &
$\left( \begin{array}{cc} 1 & 0 \\ 0 & 1 \end{array} \right)$ &
$\left( \begin{array}{cc} 1 & 0 \\ 0 & 1 \end{array} \right)$ &
$\left( \begin{array}{cc} \frac{1}{\sqrt{2}} & -\frac{1}{\sqrt{2}} \\ \frac{1}{\sqrt{2}} & \frac{1}{\sqrt{2}} \end{array} \right)$  
\\
&&&&\\[-.7em]
\hline
&&&&\\[-.7em]
$\psi =$ & 
$\widetilde{\phi}_1 \widetilde{\chi}_1$ &
$ \frac{1}{\sqrt{2}}\widetilde{\phi}_1 \widetilde{\chi}_1\newline +\frac{1}{\sqrt{2}}\widetilde{\phi}_2 \widetilde{\chi}_2$ &
$\widetilde{\phi}_1 \widetilde{\chi}_1$ &
$\widetilde{\phi}_1 \widetilde{\chi}_1$ \\
&&&&\\[-.7em]
\hline
&&&&\\[-.7em]
$\widetilde{\phi}_1= $ & 
$\phi_1$  & 
$\phi_2$  & 
$\frac{1}{\sqrt{2}}\left( \phi_1  + \phi_2  \right)$ &
$\phi_1$ \\
&&&&\\[-.7em]
\hline
&&&&\\[-.7em]
$\widetilde{\phi}_2= $ & 
---  & 
$\phi_1$  & 
--- &
--- \\
&&&&\\[-.7em]
\hline
&&&&\\[-.7em]
$\widetilde{\chi}_1= $ & 
$\chi_1$  & 
$\chi_1$  & 
$\chi_1$  & 
$\frac{1}{\sqrt{2}}\left( \chi_1  + \chi_2  \right)$  \\
&&&&\\[-.7em]
\hline
&&&&\\[-.7em]
$\widetilde{\chi}_2= $ & 
---  & 
$\chi_2$  & 
--- &
--- \\
&&&&\\[-.7em]
\hline \hline
\end{tabular}
\end{table}

In case (i), the exciton involves only a single transition from the highest occupied orbital $\phi_1$ to the lowest unoccupied orbital $\chi_1$, which is, in fact, a common case for the lowest exciton in some organic molecules~\cite{Wallauer2020}. Evidently, exPOT predicts that the observed PAD is given by the Fourier transform of $\chi_1$ appearing at the kinetic energy $E_\mathrm{kin}  = \omega - \varepsilon_1 + \Omega_1$, where $\omega$ is the probe photon energy, $\varepsilon_1$ the ionization potential corresponding to $\phi_1$, and $\Omega_1$ denotes the exciton energy, i.e. the pump photon energy. This is also illustrated in the bottom part of Figure~\ref{figure1}, where the the square above the peak in the sketched kinetic energy spectrum should represent the expected PAD map of $\chi_1$. Also note that the NTOs coincide with the original orbitals in this case. For case (ii), we assume the exciton wave function as $\psi = \frac{1}{\sqrt{2}}(\phi_2 \chi_1 + \phi_1 \chi_2)$.
Here, Eq.~\ref{eq:finalintensity} leads to PAD maps of both $\chi_1$ and $\chi_2$, however, appearing at kinetic energies corresponding to the ionization levels of $\phi_2$ and $\phi_1$, respectively, as also illustrated in Figure~\ref{figure1}. 
Note that this exciton represents a truly entangled state~\cite{Plasser2016} which can also be seen after transforming to the NTO basis (see Table~\ref{tab:fourcases}).
The situation is somewhat different for case (iii), where we assume $\psi = \frac{1}{\sqrt{2}}(\phi_2 \chi_1 + \phi_1 \chi_1)$. Here, we expect to observe two identical PADs appearing at two different kinetic energies, depending on whether, after the electron has been emitted, the hole resides in state $\phi_1$ or $\phi_2$. While the unoccupied NTO $\widetilde{\chi}_1$ equals $\chi_1$, the two occupied orbitals can now be represented by a single NTO. Finally in case (iv), the exciton is described by $\psi = \frac{1}{\sqrt{2}}(\phi_1 \chi_1 + \phi_1 \chi_2)$ and Eq.~\ref{eq:finalintensity} suggests that the PAD is proportional to the Fourier transform of a \emph{coherent} sum of the unoccupied orbitals $\chi_1$ and $\chi_2$, the NTO $\widetilde{\chi}_1$, which appears at $E_\mathrm{kin}  = \omega - \varepsilon_1 + \Omega_1$. In the following, we want to give examples for the non-trivial cases (ii)--(iv) by actual numerical simulations.

\section{Results and Discussion}
Let us now compare the predictions of our exPOT approach for organic molecules with computationally more demanding, but accurate TDDFT calculations as implemented in the real-space code OCTOPUS~\cite{Andrade2015,Tancogne-Dejean2020}. Here, photoemission spectra and PAD maps are obtained by recording the flux of photoelectron density through a detector surface (t-SURFF)~\cite{Wopperer2017,DeGiovannini2017}, which seamlessly allows for pump-probe setups and where no assumptions on the final state need to be made. 

For a better comparability of the two theoretical approaches, exPOT vs. t-SURFF, we take several precautions. 
First, we focus on planar molecules for which the plane wave approximation has already been well tested~\cite{Luftner2013,Kliuiev2019}.
Second, we choose the probe field in $z$-direction, that is perpendicular to the molecular plane, which is also known to minimize  possible deficiencies of the plane wave approximation (PWA)~\cite{Dauth2016a}. 
Third, we ensure that pump pulses are long enough to only excite the specific exciton in question, since for ultrashort pulses considerable energy broadening needs to be taken into account~\cite{Popova2016,Reuner2023}. Equivalently, we keep our probe pulses long enough for a resonable kinetic energy resolution in the spectra and choose probe energies in the XUV regime for the sake of the sudden approximation~\cite{Hammon2021}.
Fourth, we limit ourselves to the adiabatic local density approximation (ALDA) since more advanced functionals, such as hybrids, would be computationally too demanding for the real-time propagation utilized for the t-SURFF method. We emphasize, however, that for the application of our exPOT formalism, the latter restriction is not necessary and any method for excited states that provides a transition density matrix in terms of single-particle orbitals can be used.

\begin{figure}
\begin{center}
\includegraphics[width=\columnwidth]{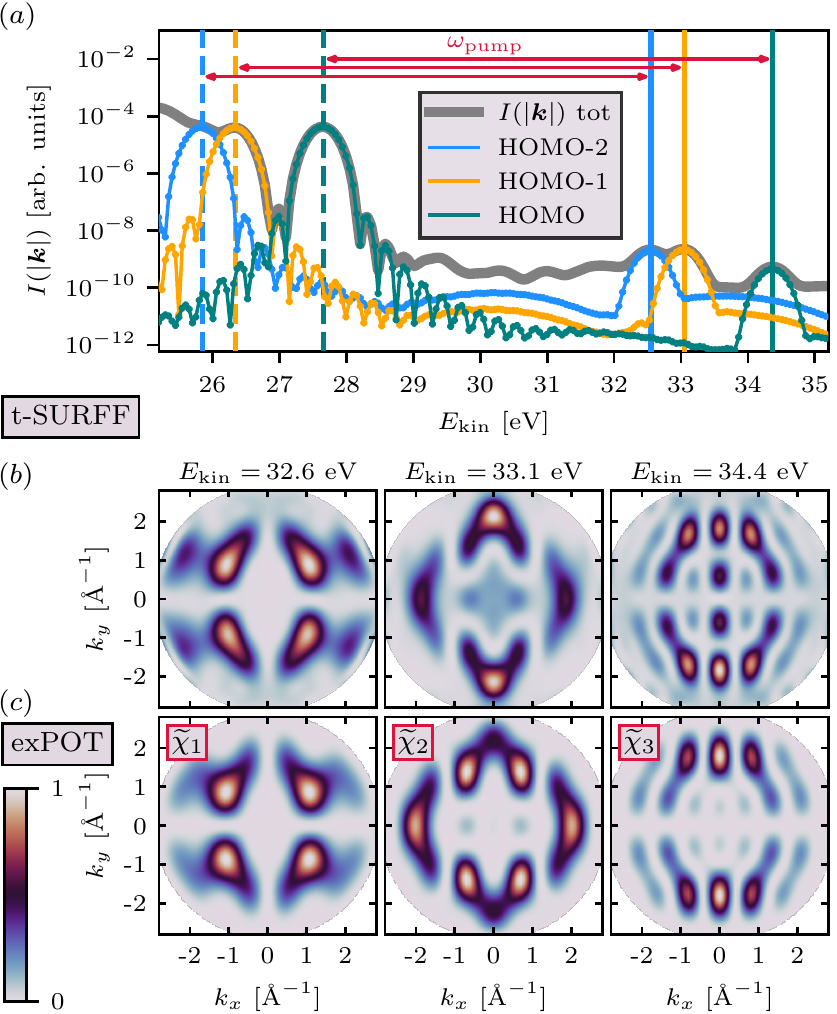}
\caption{\textbf{Comparison of exPOT with results from t-SURFF for TCNQ.} (a) total angle-integrated photoelectron intensity from t-SURFF (grey) and its projection onto the HOMO ($v$=1, green), HOMO-1 ($v$=2, orange) and HOMO-2 ($v$=3, blue) states, with corresponding kinetic energy positions $\omega - \varepsilon_j$ indicated by the vertical dashed lines in the same colors. Red arrows mark the energy of the pump pulse $\omega_{\mathrm{pump}}$, full vertical lines $\omega - \varepsilon_j + \omega_{\mathrm{pump}}$ respectively. (b) PAD maps from t-SURFF at the kinetic energies indicated by the full vertical lines in panel (a). 
(c) PAD maps  obtained from the exPOT approach with the first three NTOs.}
\label{figure2}
\end{center}
\end{figure}
With the aim to find real-life examples for the cases (ii)--(iv) outlined above, we have selected three prototypical $\pi$-conjugated molecules, namely tetracyanoquinodimethane (TCNQ), porphine and perylenetetracarboxylic dianhydride (PTCDA), and perform linear-response TDDFT calculations within the Casida formalism in OCTOPUS. The details of those calulations are described in Appendix~\ref{sec:casida}.
For TCNQ, the solution reveals an exciton with $\Omega_m=6.76$~eV 
which is strongly allowed for $y$-polarization (molecular geometry and choice of axis are depicted in Sec.~\ref{sec:casida}). Its exciton wave function has major contributions from $\phi_3 \chi_2$ (0.44), $\phi_2 \chi_3$ (0.35) and $\phi_1 \chi_6$ (0.07) (see Table~\ref{tab:casida_states} for more details). Thus it represents an entangled state as in case (ii).
In the t-SURFF calculations, we set the pump energy $\omega_\mathrm{pump} = \Omega_m$ and employ a probe energy of $\omega=35$~eV (details in Appendix~\ref{sec:tddft}). The resulting kinetic energy spectrum of the emitted electrons is depicted in panel (a) of Fig.~\ref{figure2}. It is dominated by emissions from the three highest occupied orbitals $\phi_1$, $\phi_2$ and $\phi_3$ indicated by the green, orange and blue dashed vertical lines, respectively. Importantly, however, we also observe three emission peaks at kinetic energies larger by precisely $\omega_\mathrm{pump}$. This behavior, already qualitatively illustrated in the second column of Fig.~\ref{figure1}, is in perfect accordance with the energy conservation of Eq.~\ref{eq:finalintensity}.
Despite the orders of magnitude smaller peak heights for the exciton emission, we obtain three distinct PAD maps (at the kinetic energies marked by vertical full lines), which are displayed in panel (b). Comparing with our exPOT theory, indeed, the Fourier transforms of the first three NTOs of this entangled exciton, as depicted in panel (c), are in very good agreement with the PAD maps from t-SURFF.

\begin{figure}
\begin{center}
\includegraphics[width=\columnwidth]{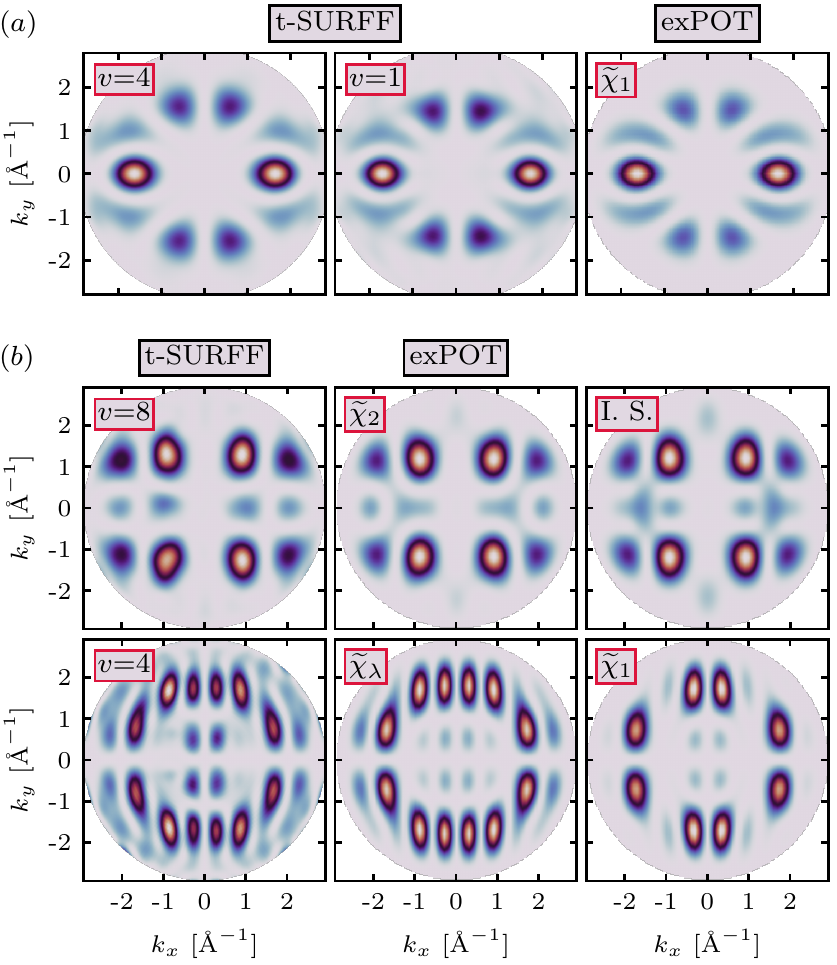}
\caption{\textbf{Comparison of exPOT with results from t-SURFF for porphin and PTCDA.} (a) PADs for porphin from t-SURFF at different kinetic energies (left and middle column) compared to the exPOT map of the first NTO (right column). (b) Different methods for PTCDA, showing contributions from $v=4$ (top row) and $v=8$ (bottom row), see text for details.}
\label{figure3}
\end{center}
\end{figure}

Next, we present our results for the optical excitation in porphin at $\Omega_m=3.94$~eV in $x$-direction, which serves as an example for case (iii) defined in Figure~\ref{figure1}. 
From the t-SURFF calculation, we obtain two identical momentum maps at the kinetic energies corresponding to the hole in state $\phi_1$ and $\phi_4$ (left and middle column of panel (a) in Fig.~\ref{figure3}). Note that here, in contrast to the above PADs from TCNQ, we have projected the t-SURFF ARPES intensities on the respective ground-state orbitals, since the total photoelectron yield is also affected by other contributions which are not relevant for our case (see also Sec.~\ref{sec:tddft}). 
The Casida calculation leads to almost equal contributions of $\phi_1 \chi_2$ (0.27) and $\phi_4 \chi_2$ (0.25) to the exciton wave function, which can be written as a single NTO $\widetilde{\chi}_1$ (see Table~\ref{tab:casida_states}), resulting in the PAD depicted in the rightmost column of panel (a) in Fig.~\ref{figure3}.
The excellent agreement with the corresponding t-SURFF maps further validates the exPOT predictions. Remarkably, while a single NTO might be enough to explain photoemission from an excited state of such character, it can be comprised of contributions from different valence states, which then lead to photoemission signatures of the same conduction state at different kinetic energies. 

Conversely, in case (iv), we consider an exciton structure with transitions involving only a single hole state $v$ but multiple conduction states $c$.
For PTCDA at an excitation energy of $\Omega_m=4.45$ eV (polarized in $y$-direction), we encounter even two such transitions: $\phi_8 \chi_4$ (0.29), $\phi_8 \chi_3$ (0.03) and $\phi_4 \chi_8$ (0.06), $\phi_4 \chi_2$ (0.06). 
The top row of panel (b) of Fig.~\ref{figure3} is devoted to the contribution from $v=8$, with the state-projected result from t-SURFF in the left column agreeing very well with the exPOT result in the middle column, evaluated with the contribution from $\widetilde{\chi}_2$ only. Importantly, the computation of the latter implicitly involves a \emph{coherent} sum over the unoccupied states $\chi_4$ and $\chi_3$, while wrongly performing an \emph{incoherent} summation worsens the agreement for with the t-SURFF reference (see right panel labeled I. S.).
The second major set of contributions to this exciton, $\phi_4 \chi_8$ and $\phi_4 \chi_2$, leads to a PAD at the kinetic energy corresponding to $\varepsilon_4$ and is shown in the bottom row of Fig.~\ref{figure3}, panel (b). Again, the t-SURRF result (left column) agrees well with exPOT (middle column). This time however, we need to take into account a sum over multiple NTOs ($\widetilde{\chi}_{\lambda}$) while the PAD from a single NTO ($\widetilde{\chi}_1$, right column) is not sufficiently accurate. This is due to the fact that, in general, the electron or hole contributions can contribute to different NTOs and only the coherent sum over $\lambda$ is equivalent to the coherent sum of Eq.~\ref{eq:finalintensity} (see also comparison of PADs in Appendix~\ref{sec:com_results}). In summary, we have not only proven excellent agreement of the exPOT theory with \emph{ab-initio} simulations for case (iv), but could also emphasize the necessity of the \emph{coherent} superposition of the electron orbitals for such a case.

\section{Conclusions}
We demonstrate an extension of photoemission orbital tomography to excitons, termed exPOT, and thereby provide the theoretical foundations to interpret photoemission angular distributions maps as measured in pump-probe ARPES experiments of oriented organic molecules in terms of exciton wave functions. 
We illustrate the consequences of exPOT on the example of three  organic molecules, covering a range of prototypical exciton structures, and validate our findings by real-time TDDFT calculations that directly incorporate the pump and probe fields. 
In our method, the simplicity of the orbital interpretation can be retained by identifying Fourier-transformed NTOs as the observables in photoemission of excitons. The evaluation of the ARPES intensity, however, demands a coherent sum over electron contributions to reflect the entangled character of an exciton wave function, as well as an incoherent sum over hole contributions to fulfill energy conservation.
While in this work, we have restricted ourselves to organic molecules in the gas phase, the extension of exPOT to periodic systems and magnetic materials is straight-forward. Moreover, our method can also be combined with any common excited state description, e.g.~including electron-hole correlations within the framework of the Bethe-Salpeter equation.

\begin{acknowledgments}
The authors thank Wiebke Bennecke, G. S. Matthijs Jansen and Stefan Mathias for valuable discussions. This work was supported by the Austrian  Science Fund project I~4145, the Doctoral Academy NanoGraz and from the European Research Council (ERC) Synergy Grant, project ID~101071259. We further acknowledge computational resources at the Vienna Scientific Cluster.
\end{acknowledgments}

\appendix
\section{Ground state and linear response calculations}
\label{sec:casida}
\begin{figure}
\begin{center}
\includegraphics[width=4.5cm]{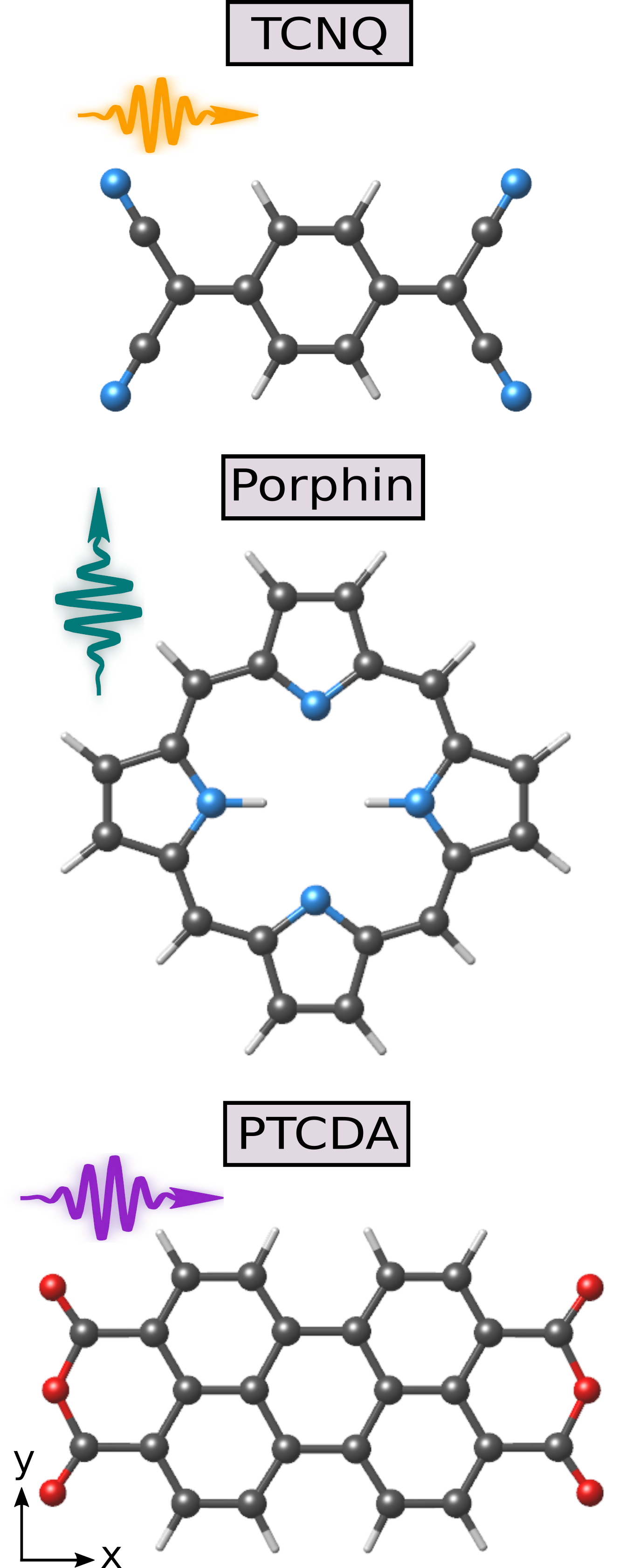}
\caption{Geometries of the three molecules used in our investigation (TCNQ, porphin and PTCDA), arrows mark the incident direction of the pump pulse in the photoemission simulation}
\label{figure4.png}
\end{center}
\end{figure}

The structures of the three molecules TCNQ (C$_{12}$H$_{4}$N$_{4}$), porphin (C$_{20}$H$_{14}$N$_{4}$) and PTCDA (C$_{24}$H$_{8}$O$_{6}$) were optimized using the real-space mode of GPAW~\cite{Mortensen2005,Enkovaara2010} in conjunction with the BFGS minimization routine from the Atomic Simulation Environment (ASE)~\cite{Larsen2017}. We used a simulation box with $0.2$~$\mathrm{\AA}$ spacing, $8$~$\mathrm{\AA}$ vacuum around each molecule and set the maximum force criterion to $0.02$~eV$/\mathrm{\AA}$. These relaxed geometries were then used in all further calculations and are depicted in Fig.~\ref{figure4.png} together with the Cartesian coordinate system and the direction of the pump field incidence.

In order to solve Casida's equation and perform the NTO analysis, we employed the linear-response TDDFT (LR-TDDFT) implementation of the real-space code OCTOPUS~\cite{Andrade2015,Tancogne-Dejean2020}.
For the three molecules, we used a simulation domain with spheres of radius $8$~$\mathrm \AA$ around each atom and a spacing of $0.2$~$\mathrm\AA$. While the latter value for the spacing may not lead to fully converged results for the geometry optimization described before, as well as for the optical spectra in the following, we choose $0.2$~$\mathrm \AA$ none the less for all calculations to be consistent with the numerically very demanding ARPES simulations. For the same reason, we used the local density approximation (LDA)~\cite{Dirac1930} for LR-TDDFT calculations with the Perdew-Zunger implementation of correlation~\cite{Perdew1981} and norm-conserving Troullier-Martins pseudopotentials~\cite{Troullier1991}. Having computed the respective groundstate of the three molecules this way, we solved Casida's equation with the same numerical parameters and considered an energy window of $32$~eV, $28$~eV and $30$~eV for TCNQ, Porphin and PTCDA, respectively. In this range combinations of occupied and unoccupied states were considered, thereby obtaining the transition density matrices $X_{v c}^{(m)}$ for the $m$-th exciton. Note that our calculations also include de-excitations beyond the Tamm-Dancoff approximation. For the NTOs, we computed the singular value decomposition of Eq.~\ref{eq:svd} with python's numpy package~\cite{Harris2020}.
The results of the LR-TDDFT calculations are shown in Table~\ref{tab:casida_states} and optical spectra are shown in Fig~\ref{figure_spectra} for comparison with the real-time TDDFT calculations of the next section.
\begin{table}
\caption{Casida excitation energies, $\Omega_m$, their corresponding single particle contributions in terms of the inital Kohn-Sham molecular orbitals, $X_{v c}^2$, and the eigenvalues (magnitudes) of the natural transition orbitals, $\Lambda_{\lambda}^2$, for the three molecules presented in the main text. All contributions greater than $0.01$ are shown, those referenced in our investigation are highlighted in color, where the different shadings connect the single-particle contributions with their NTO counterparts, with the exception of the NTOs for PTCDA, since here a full sum over $\lambda$ was necessary (see text for details)}. {\label{tab:casida_states}}
\begin{tabular}{cccccccc}
\hline \hline
&&&&&&&\\[-.7em]
\multicolumn{2}{c}{TCNQ} & & \multicolumn{2}{c}{porphin} & & \multicolumn{2}{c}{PTCDA} \\
&&&&&&&\\[-1em]
\multicolumn{2}{c}{$\Omega_m=6.76~\mathrm{eV}$} & & \multicolumn{2}{c}{$\Omega_m=3.52~\mathrm{eV}$} & & \multicolumn{2}{c}{$\Omega_m=5.51~\mathrm{eV}$} \\
&&&&&&&\\[-.7em]
\hline
&&&&&&&\\[-.7em]
$ \phi_v \rightarrow \chi_c $ & $X_{vc}^2$ & & $ \phi_v \rightarrow \chi_c $ & $X_{vc}^2$ & & $ \phi_v \rightarrow \chi_c $ & $X_{vc}^2$ \\
\cellcolor{orange!75}$\boldsymbol{3 \rightarrow 2}$ & $\boldsymbol{0.44}$ & & 
$2 \rightarrow 1$ & $0.36$ & & \cellcolor{violet!50}$\boldsymbol{8 \rightarrow 4}$ & $\boldsymbol{0.29}$ \\
\cellcolor{orange!50}$\boldsymbol{2 \rightarrow 3}$ & $\boldsymbol{0.35}$ & & 
\cellcolor{teal!50}$\boldsymbol{1 \rightarrow 2}$ & $\boldsymbol{0.27}$ & & $ 11 \rightarrow 2$ & $0.23$ \\
\cellcolor{orange!25}$\boldsymbol{1 \rightarrow 6}$ & $\boldsymbol{0.07}$ & & 
\cellcolor{teal!50}$\boldsymbol{4 \rightarrow 2}$ & $\boldsymbol{0.25}$ & & $ 7 \rightarrow 7 $ & $ 0.17 $ \\
$ 5 \rightarrow 4 $ & $ 0.03 $ & & \cellcolor{teal!50}$\boldsymbol{8 \rightarrow 2}$ & $\boldsymbol{0.05}$ & & \cellcolor{violet!25}$\boldsymbol{4 \rightarrow 8}$ & $\boldsymbol{0.06}$ \\
${17 \rightarrow 2}$ & $0.02$         & & $ 3 \rightarrow 3 $ & $ 0.04 $ & & \cellcolor{violet!25} $\boldsymbol{4 \rightarrow 2}$ & $ \boldsymbol{0.06} $ \\
${11 \rightarrow 3}$ & $0.02$ & &                     &          & & $ 1 \rightarrow 5 $ & $ 0.05 $ \\
                    &          & &                     &          & & \cellcolor{violet!50}$\boldsymbol{8 \rightarrow 3}$ & $\boldsymbol{0.03}$ \\
                    &          & &                     &          & & ${16 \rightarrow 1}$ & $0.02$ \\
                    &          & &                     &          & & ${7 \rightarrow 1}$ & $0.02$ \\
                    &          & &                     &          & & ${9 \rightarrow 2}$ & $0.02$ \\
&&&&&&&\\[-.7em]
\hline 
&&&&&&&\\[-.7em]
$ \widetilde{\phi}_{\lambda} \rightarrow \widetilde{\chi}_{\lambda} $ & $\Lambda_{{\lambda}}^2$ & & $ \widetilde{\phi}_{\lambda} \rightarrow \widetilde{\chi}_{\lambda} $ & $\Lambda_{{\lambda}}^2$ & & $ \widetilde{\phi}_{\lambda} \rightarrow \widetilde{\chi}_{\lambda} $ & $\Lambda_{\lambda}^2$ \\
\cellcolor{orange!75}$\boldsymbol{1 \rightarrow 1}$ & $\boldsymbol{0.46}$ & & 
\cellcolor{teal!50}$\boldsymbol{1 \rightarrow 1}$ & $\boldsymbol{0.57}$ & & 
$\boldsymbol{1 \rightarrow 1}$ & $\boldsymbol{0.32}$ \\
\cellcolor{orange!50}$\boldsymbol{2 \rightarrow 2}$ & $\boldsymbol{0.39}$ & & 
$ 2 \rightarrow 2 $ & $ 0.36 $ & & $\boldsymbol{2 \rightarrow 2}$ & $\boldsymbol{0.32}$ \\
\cellcolor{orange!25}$\boldsymbol{3 \rightarrow 3}$ & $\boldsymbol{0.07}$ & & 
$ 3 \rightarrow 3 $ & $ 0.05 $ & & $ 3 \rightarrow 3 $ & $ 0.20 $ \\
$ 4 \rightarrow 4 $ & $ 0.04 $ & &                     &          & & $ 4 \rightarrow 4 $ & $ 0.06 $ \\
$ 5 \rightarrow 5 $ & $0.01$ & &                     &          & & $ 5 \rightarrow 5 $ & $ 0.05 $ \\
                    &          & &                     &          & & $ 6 \rightarrow 6 $ & $ 0.03 $ \\
&&&&&&&\\[-.7em]
\hline \hline
\end{tabular}
\end{table}

\section{Real-time TDDFT calculations}
\label{sec:tddft}
\begin{figure*}
\begin{center}
\includegraphics[width=11cm]{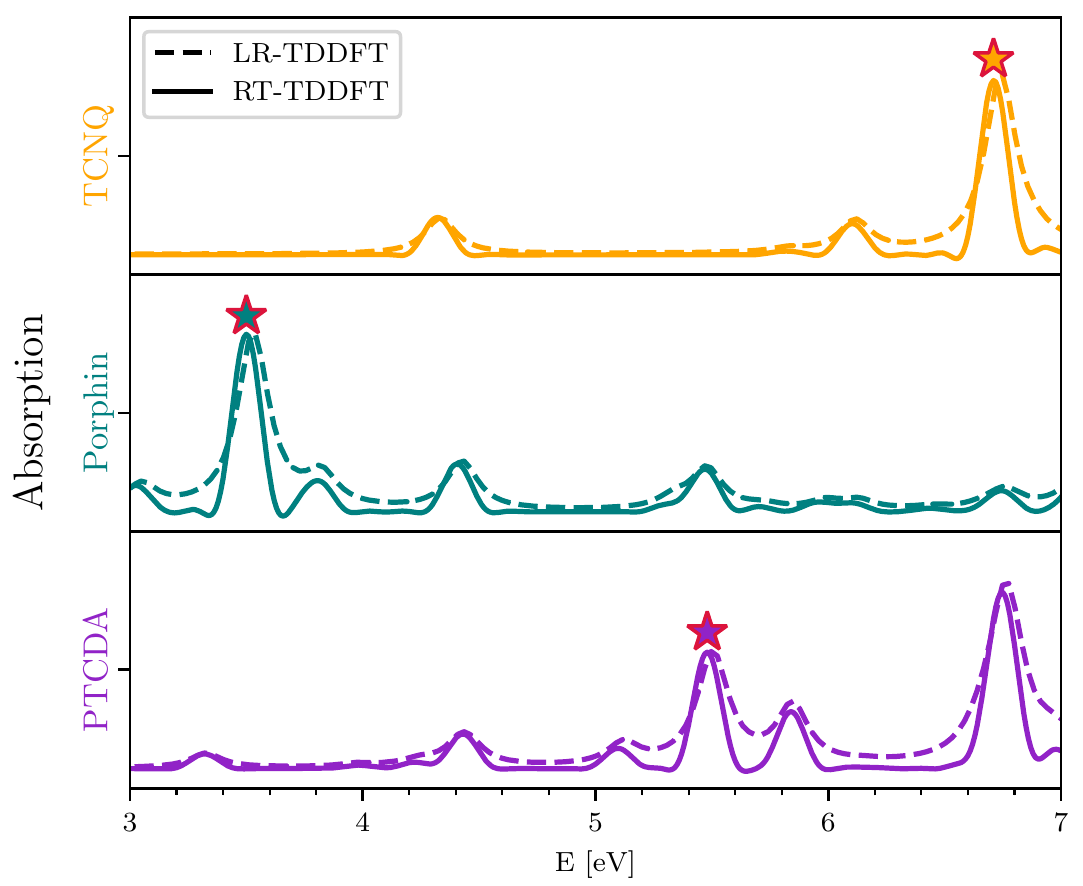}
\caption{Absorption spectra of the molecules TCNQ, porphin and PTCDA calculated with OCTOPUS in RT-TDDFT (full curves) and within the linear-response Casida formalism (dashed curves). Excitation energies used in the pump-probe photoemission simulations are marked with an asterisk.}
\label{figure_spectra}
\end{center}
\end{figure*}

In this section, we describe the methods to obtain the ab-initio simulations of photoemission from real-time TDDFT (RT-TDDFT) with OCTOPUS.
While in the last section, the results for linear-response calculations already delivered the desired excitation energies, we also employed a RT-TDDFT method for optical spectra~\cite{Yabana1996}. Using the ground state calculations with the same parameters as described in the previous section, we perturbed the system at initial time $t=0$ with a Dirac-$\delta$ pulse (pulse strength: 0.01~$\mathrm{\AA}^{-1}$) that equally excites all optically allowed transitions. We then evolved the system for further $30$~fs, with a time steps of 2~as, and Fourier transformed the time-dependent dipole-moment to get the optical spectrum~\cite{Yabana2006}. In Fig.~\ref{figure_spectra}, we compare the optical spectra from RT-TDDFT with those from the LR-TDDFT calculations of the previous section. For all three molecules, we find very good agreement, thus assuring the comparability of our methods. Since we also use TDDFT in the real-time fashion for the ARPES simulations, we use the excitation energies (marked by $\star$ symbols) from RT-TDDFT.

Having obtained the excitation energies of interest, we now describe the method used for the pump-probe ARPES simulations with t-SURFF~\cite{Wopperer2017,DeGiovannini2017}. For all three molecules, we first computed the ground state as described above, with the only difference that we used a spherical simulation box around the center of the molecules with $R = 35$~$\mathrm \AA$ radius. Then, the systems were subjected to pump pulses with respective energies $\Omega_m$ for $t_{\mathrm{pump}}=20$~fs, followed by $t_{\mathrm{probe}}=15$~fs of propagation time with the probe pulse. While the energy and direction of the pump pulses were varied according to the excitations within the different molecules, we always probed with $z$-polarized fields and a photon energy of $\omega = 35$~eV. For both types of pulses, we used a $\cos(\omega t)$ function, shaped by a hull function of $\sin^2$-type to ensure gradual on- and off-switching of the fields, thereby avoiding non-resonant excitations. The field amplitudes were varied such that the radiation would correspond to a laser with intensity $10^8$~W$/$cm$^2$. In order to avoid spurious effects of reflected electron density at the border of our simulation region, we inserted a complex absorbing potential (CAP)~\cite{DeGiovannini2015} described by $\mathrm i \xi \sin^2(\frac{\Theta(r-R_0) \pi}{2 R})$, with magnitude $\xi=-0.2$~a.u. and onset at $R_0=20$~$\mathrm \AA$. Over all times, we recorded the flux of electron density through a spherical surface~\cite{Wopperer2017,DeGiovannini2017} at $R_0$ and thus obtained energy- and angle-resolved photoemission intesities in an \emph{ab-inito} way as a direct numerical simulation of the experiment.

\section{Complementing results}
\label{sec:com_results}
\begin{figure}
\begin{center}
\includegraphics[width=11cm]{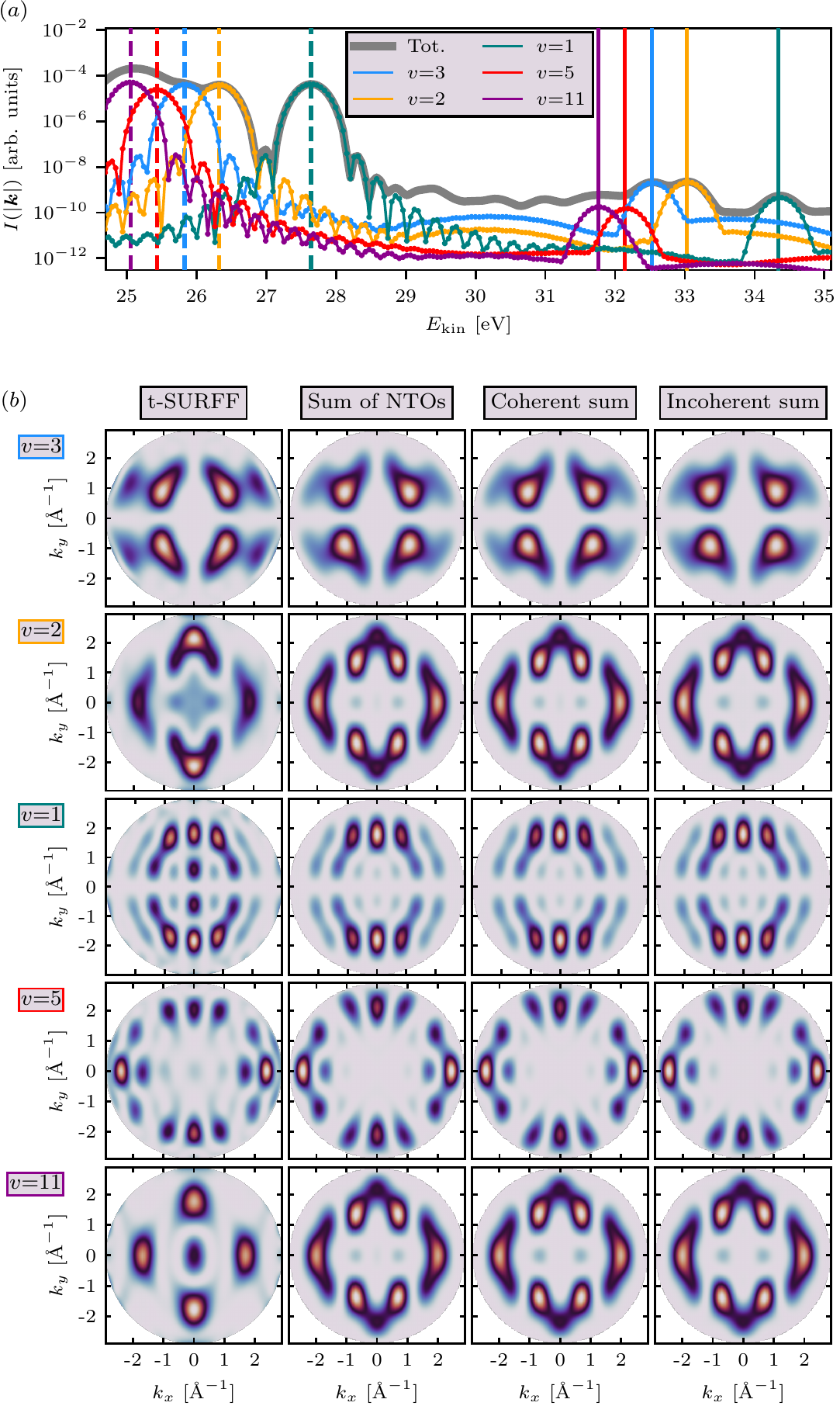}
\caption{\textbf{Summary of results for TCNQ excited with 6.7 eV in $\boldsymbol y$-direction.} The kinetic energy spectrum from t-SURFF is shown in panel (a) with $I(|\boldsymbol k|)$ in grey, as well as the projection on states $v=\{3,2,1,5,11\}$. In the same colors, we show $E_v$ in dashed lines and $E_v + \omega_{\mathrm{pump}}$ in full lines. In panel (b), the corresponding momentum maps of the state-projected photoemission intensities from t-SURFF are shown in each line of the leftmost column. In the left-middle column, we show the results from exPOT for the sum over NTOs (Eq.~\ref{eq:nto_intensity}) and the equal results from exPOT with the coherent sum over $X_{v c} \chi_c$ (Eq.~\ref{eq:finalintensity}) in the middle-right column. For comparison, the results with a wrongly performed incoherent sum are shown in the rightmost column (see text for details).}
\label{fig:TCNQ}
\end{center}
\end{figure}

\begin{figure}
\begin{center}
\includegraphics[width=11cm]{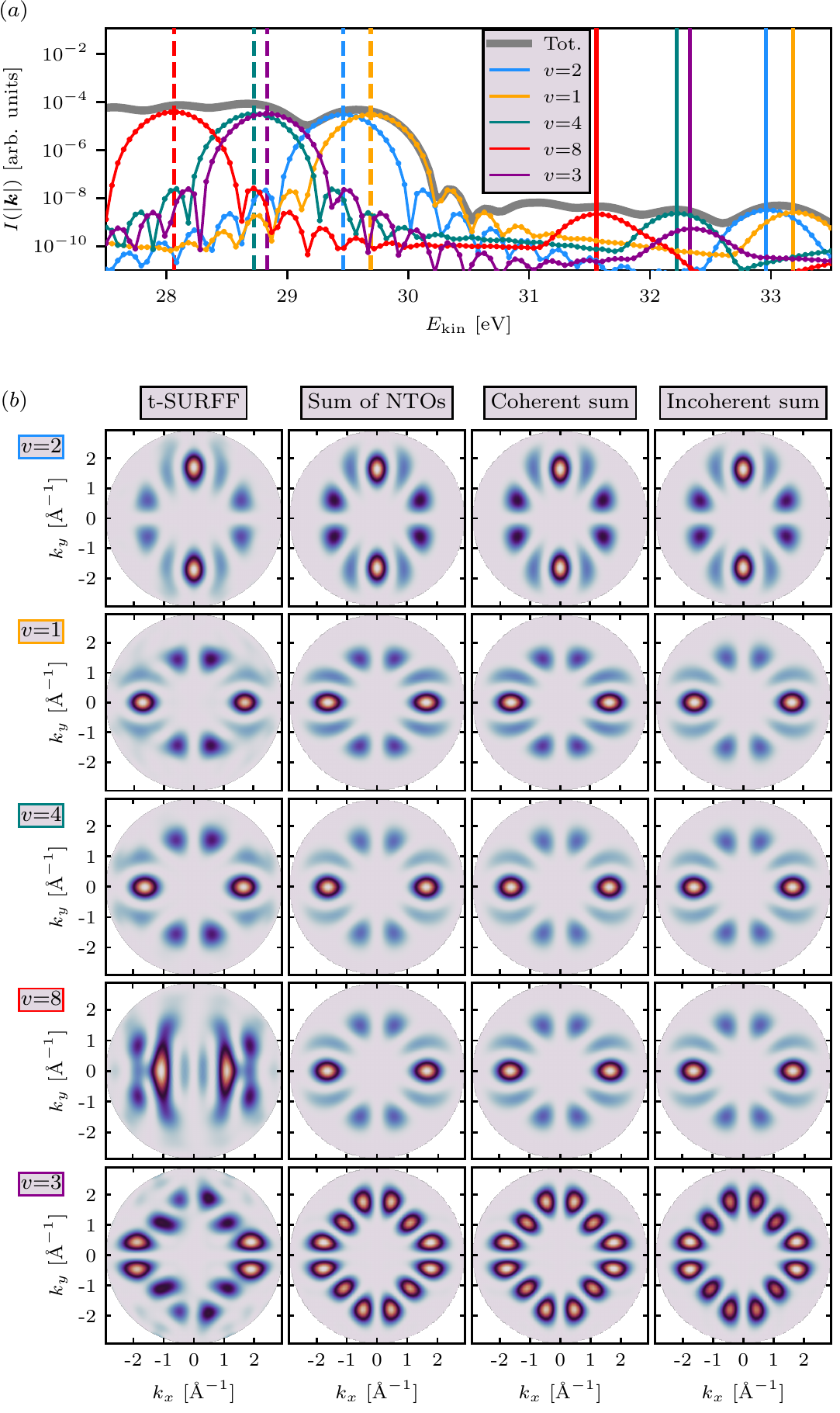}
\caption{\textbf{Summary of results for porphin excited with 3.5 eV in $\boldsymbol x$-direction.} The kinetic energy spectrum from t-SURFF is shown in panel (a) with $I(|\boldsymbol k|)$ in grey, as well as the projection on states $v=\{2,1,4,8,3\}$. In the same colors, we show $E_v$ in dashed lines and $E_v + \omega_{\mathrm{pump}}$ in full lines. In panel (b), the corresponding momentum maps of the state-projected photoemission intensities from t-SURFF are shown in each line of the leftmost column. In the left-middle column, we show the results from exPOT for the sum over NTOs (Eq.~\ref{eq:nto_intensity}) and the equal results from exPOT with the coherent sum over $X_{v c} \chi_c$ (Eq.~\ref{eq:finalintensity}) in the middle-right column. For comparison, the results with a wrongly performed incoherent sum are shown in the rightmost column (see text for details).}
\label{fig:Por}
\end{center}
\end{figure}

\begin{figure}
\begin{center}
\includegraphics[width=11cm]{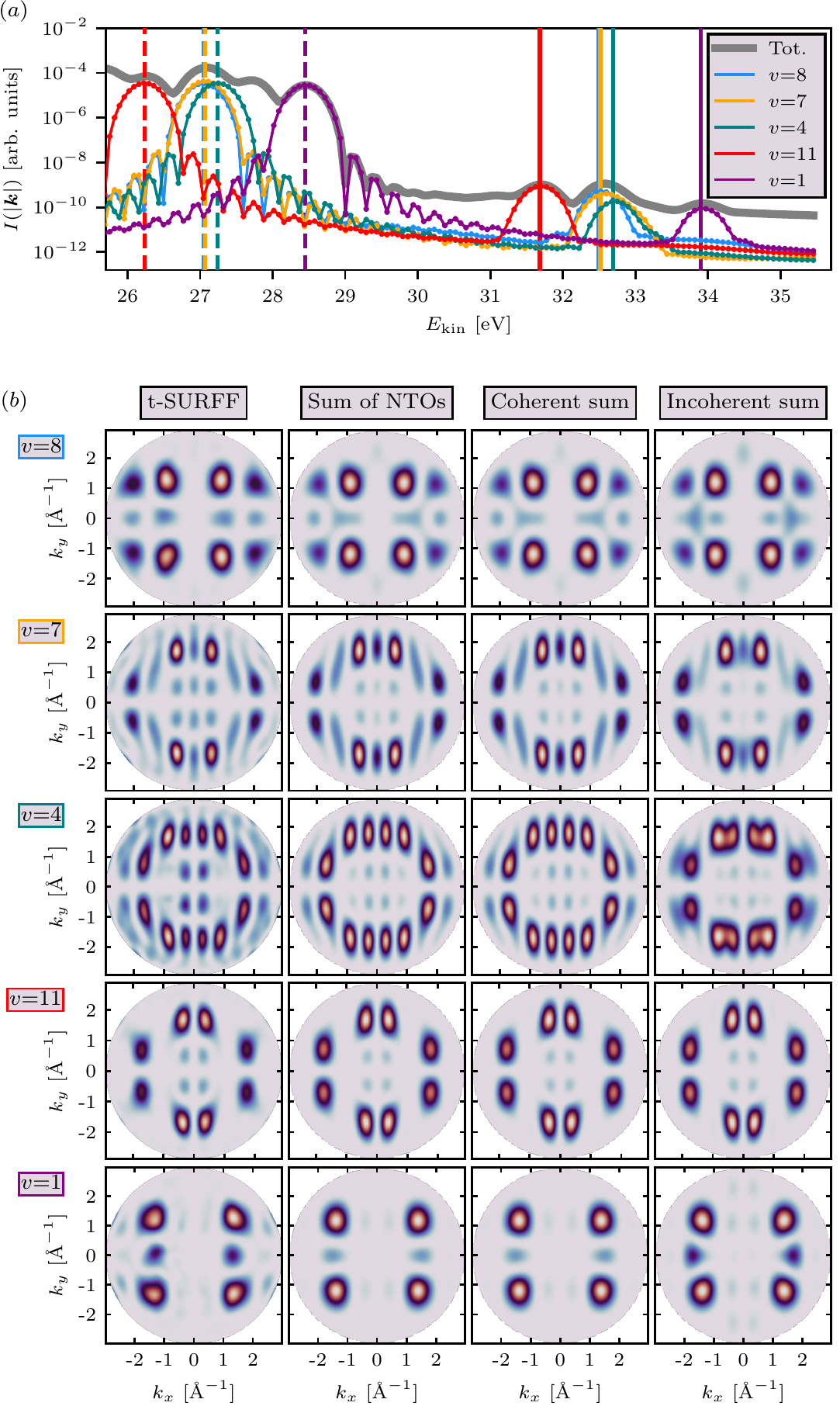}
\caption{\textbf{Summary of results for PTCDA excited with 5.45 eV in $\boldsymbol y$-direction.} The kinetic energy spectrum from t-SURFF is shown in panel (a) with $I(|\boldsymbol k|)$ in grey, as well as the projection on states $v=\{8,7,4,11,1\}$. In the same colors, we show $E_v$ in dashed lines and $E_v + \omega_{\mathrm{pump}}$ in full lines. In panel (b), the corresponding momentum maps of the state-projected photoemission intensities from t-SURFF are shown in each line of the leftmost column. In the left-middle column, we show the results from exPOT for the sum over NTOs (Eq.~\ref{eq:nto_intensity}) and the equal results from exPOT with the coherent sum over $X_{v c} \chi_c$ (Eq.~\ref{eq:finalintensity}) in the middle-right column. For comparison, the results with a wrongly performed incoherent sum are shown in the rightmost column (see text for details).}
\label{fig:PTCDA}
\end{center}
\end{figure}

In the following, we give additional results that complement those of the main text for all three molecules. For each molecule in Fig.~\ref{fig:TCNQ}--\ref{fig:PTCDA}, we show the kinetic energy spectra from t-SURFF (Panels (a)) in conjunction with momentum maps from the different methods presented for a series of orbitals that are relevant for the respective excitons (Panels (b)). For TCNQ in Fig.~\ref{fig:TCNQ}, all results between the different theoretical descriptions agree well, with the exception of maps for $v=11$, where the results from t-SURFF are different to exPOT. Interestingly, it seems that the t-SURFF map for $v=11$ depicts what seems to be missing for the exPOT map for $v=2$, i.e. the accentuation of the main feature at $k_x=0$ $\mathrm{\AA^{-1}}$, $k_y\ge2$ $\mathrm{\AA^{-1}}$.
The additional results for porphin in Fig.~\ref{fig:Por} show very good agreement as well, with the one exception of $v=8$, which does not agree at all. For the two pathological cases, $v=11$ in TCNQ and $v=8$ in porphin, we wish to remark that for both cases the contributions to the transition matrix are alread quite small (1-2 \%) such that better converged LR-TDDFT calculations might give other results. The same argument is valid for the t-SURFF calculations, where it can be seen in the kinetic energy-resolved spectra that the peaks stemming from these two transitions are by approximately an order of magnitude smaller than those of the main contributions and would hardly be detectable in an actual experiment.

\onecolumngrid

\end{document}